\documentclass[12pt]{article}
\usepackage{psfig}

\begin{document}

\begin{titlepage}
\begin{center}

{\Large\bf{Quarkonium suppression as a probe of a saturated  gluon plasma?}}
\\[5.0ex]
{ \Large \it{ V. P.  Gon\c{c}alves
$^{*}$\footnotetext{$^{*}$E-mail:barros@ufpel.tche.br;
barros@if.ufrgs.br} }}
\\[1.5ex]
 {\it  Instituto de F\'{\i}sica e Matem\'atica, Univ. Federal
de Pelotas}\\ {\it Caixa Postal 354, 96010-090 Pelotas, RS, BRAZIL}\\[5.0ex]
\end{center}

{\large \bf Abstract:}

A dense parton system is expected to be formed in the early stage
of relativistic heavy-ion collisions at RHIC energies and above.
The probability of a quark gluon plasma production and the
resulting strength of its signatures depends strongly on the
initial conditions associated to the distributions of partons in
the nuclear wave functions. At very high energies, the growth of
parton distributions should saturate, possibly forming a Color
Glass Condensate, which is characterized by a bulk momentum scale
$Q_s$. As a direct consequence, the possible signatures of the QGP
should be $Q_s$-dependent if the saturation scenario is valid for
RHIC and LHC.  In this Letter we assume  the saturation scenario
for the QGP formation and estimate the saturation scale dependence
of  quarkonium suppression. We conclude that, if this scenario is
valid, the $\Upsilon$ suppression only occurs at large values of
$Q_s$.

\vspace{1.5cm}



\end{titlepage}

High energy heavy-ion collisions offer the opportunity to study
the properties of the predicted QCD phase transition to a locally
deconfined quark-gluon plasma (QGP) \cite{rep}. A dense parton
system is expected to be formed in the early stage of relativistic
heavy-ion collisions at RHIC (Relativistic Heavy Ion Collider)
energies and above, due to onset of hard and semihard parton
scatterings. These processes happen in a very short time scale
after which a dense parton gas will be formed, not immediately in
thermal and chemical equilibrium. Secondary interactions among the
produced partons may lead to thermalization if the interactions
are sufficiently strong, allowing to study the subsequent
dynamical evolution of the system using hydrodynamic-based models
\cite{bjo}. Exactly how this parton gas equilibrates and a
thermalized plasma of quarks and gluons is formed requires solving
a kinetic equation, given appropriate initial conditions and a
detailed knowledge of the microscopic processes by which the
quanta exchange energy and momentum. This topic  is currently
under intense debate \cite{bjoraker,nayak,serreau}.

 The search for experimental evidence of this transition  during
the very early stage of the $AA$ reactions requires to extract
unambiguous characteristic signals that survive the complex
evolution through the later stages of the collision. One of the
proposed signatures of the QCD phase transition is the suppression
of quarkonium production, particularly of the $J/\psi $
\cite{satz}. The idea of suppression of $c\overline{c}$ mesons
$J/\psi $, $\psi ^{,}$, etc., is based on the notion that
$c\overline{c}$ are produced mainly via primary hard collisions of
energetic gluons during the preequilibrium stage up to shortly
after the plasma formation (before the initial temperature drops
below the production threshold), and the mesons formed from these
pairs may subsequently experience deconfinement when traversing
the region
of the plasma. In a QGP, the suppression occurs due to the shielding of the $%
c\overline{c}$ binding potential by color screening, leading to
the breakup
of the resonance. The $c\overline{c}$ ($J/\psi ,\psi ^{\prime },...$) and $b%
\overline{b}$ ($\Upsilon ,\Upsilon ^{\prime },...$) resonances have smaller
radii than light-quark hadrons and therefore higher temperatures are needed
to dissociate these quarkonium states.

The probability of a thermalized QGP production and the resulting
strength of its signatures strongly depends  on the initial
conditions associated to the distributions of partons in the
nuclear wave functions. At very high energies, the growth of
parton distributions should saturate, possibly forming a Color
Glass Condensate \cite{iancu} [For a pedagogical presentation  see
Ref. \cite{mcllec}], which is characterized by a bulk momentum
scale $Q_s$. This regime is characterized by the limitation on the
maximum phase-space parton density that can be reached in the
hadron wave function (parton saturation) and very high values of
the QCD field strength squared $F^2_{\mu \nu} \propto 1/\alpha_s$
\cite{muesat}. In this case, the number of gluons per unit phase
space volume practically saturates and at large densities grows
only very slowly (logarithmically) as a function of the energy
\cite{vicsat}. If the saturation scale is larger than the QCD
scale $\Lambda _{QCD}$, then this system can be studied using weak
coupling methods.  The magnitude of $Q_s$ is associated to the
behavior of the gluon distribution at high energies, and some
estimates has been obtained. In general, the predictions are
$Q_s\sim 1$ GeV at RHIC and $Q_s\sim 2-3$ GeV at LHC
\cite{gyusat,vicslope}.

Recently, Mueller has proposed idealized initial conditions for a
saturated nuclear wave function and  shown that the evolution of
single particle distributions could be described by a nonlinear
Landau equation \cite{mueevo}. This equation  was numerically
solved in Ref. \cite{bjoraker} and in a "self-consistent"
relaxation time approximation in Ref. \cite{serreau}. One of the
main conclusions of these studies is that the equilibration time
and the initial temperature of the plasma have a strong functional
dependence on the initial gluon saturation scale $Q_s$. As a
direct consequence, the possible signatures of this saturated
gluon plasma should be $Q_s$-dependent if the saturation scenario
is valid for RHIC and LHC.  We denote saturated gluon plasma, the
plasma formed from the collision of two nuclei with saturated
nuclear wave functions. Our goal in this Letter is assuming the
saturation scenario for the QGP formation,  estimate  the
quarkonium suppression and its dependence in the saturation scale
at LHC energies.

 Lets start from a
brief review of the quarkonium suppression in heavy-ion
collisions. The heavy quark pair leading to the quarkonia are
produced in such collisions on a very short time-scale $\sim
1/2m_Q$, where $m_Q$ is the mass of the heavy quark. The pair
develops into the physical resonance over a formation time $\tau
_f$ . For a confined surrounding, the bound states of
$c\overline{c}$ and $b\overline{b}$ interacting via two forces: a
linear confining potential and a color-Coulomb interaction. In the
plasma phase, the linear potential is absent due to the high
temperature leading to deconfinement and the color charge is Debye
screened by a cloud of surrounding quark-antiquarks pairs which
weaken the binding force between the $Q\overline{Q}$ pair, thus
reducing the color charge seen by the other (anti)quark.
Basically, above the critical temperature $T_c$, we have
\[
V(r)=-\frac \alpha r+\sigma r\Rightarrow V(r)=-\frac \alpha r\exp (-\mu _Dr)
\]
where $\alpha $ and $\sigma $ (the string tension) are
phenomenological parameters and $\mu _D$ is the Debye mass. As
discussed in Ref. \cite{karsch}, for values of the screening mass
above a certain critical value $\mu _i^{diss}$ ($i=J/\psi ,\psi
^{\prime },\chi _c,\Upsilon ,\Upsilon ^{\prime },\chi _b)$, the
screening become strong enough for binding to be impossible and
the resonance no longer forms in the plasma. As a result, the $Q$
and $\overline{Q}$ drift away from each other, after leaving the
interacting system to form a charm and bottom mesons, which
subsequently decays into leptons to be detected. The critical
screening mass for quarkonia are summarized in Table 1.

The discussion above implies a procedure to determine if
quarkonium suppression is expected. Basically, if we known the
evolution of the screening mass during the collision process we
can estimate what quarkonia will be effectively suppressed. At
screening mass values greater than $\mu _i^{diss}$ for values of
time above (or equal) to the formation time of the bound state
($\tau _F$),  the bound state does not be formed. Therefore, we
can analyze the quarkonium suppression considering the dynamical
evolution of the screening mass \cite{satzdin}. An expression
currently found in the literature for the screening mass is the
following:
\begin{eqnarray}
\mu ^2=-\frac{3\alpha _s}{\pi ^2}\lim_{|\overrightarrow{q}|\rightarrow
0}\int d^3k\frac{|\overrightarrow{k}|}{\overrightarrow{q}.\overrightarrow{k}}%
\overrightarrow{q}.\nabla
_{\overrightarrow{k}}f(\overrightarrow{k})\,\,,
 \label{mubiro}
\end{eqnarray}
 where $\alpha _s$ is the strong coupling constant and
$f(\overrightarrow{k})$ is the phase space density of gluons. This
expression was derived in Ref. \cite{biro}, following the standard
calculation (in the Coulomb gauge) of screening in the time-like
gluon propagator in a medium of gluon excitations. In the case of
an ideal equilibrium gluon gas, $f(\overrightarrow{k})$ is the
Bose-Einstein distribution and the above expression implies the
well-known screening mass $\mu ^2=4\pi \alpha _sT^2$. Although the
extrapolation from this expression for a system of highly
non-equilibrated gluons produced in the early stage of heavy ion
collisions be useful in the literature, is  but may not be well
defined.

An alternative expression for the screening mass has been derived
in Ref. \cite{mueevo} considering the saturation scenario for the
QGP. This scenario  is based on the observation that at very high
energies the parton densities in large nuclei reach saturation and
the number of partons becomes very large, allowing a description
by a classical effective field theory (EFT) \cite{mcl}. In this
approach (the McLerran-Venugopalan approach), the valence quarks
in the  nuclei are treated as classical sources of the color
charges, with the gluon distribution for very high nuclei obtained
by solving the classical Yang-Mills equation. The behavior of the
gluon distribution is characterized by the saturation momentum
$Q_s$, below which the distribution reach their maximum value,
i.e., the distribution exhibits saturation. When quantum
corrections are included, a Wilsonian renormalization group
picture emerges \cite{jamal}. The structure of classical fields
remains intact, the scale $Q_s$ grows because hard gluons (harder
than the $x$ scale of interest) in the source increases the
typical size of color fluctuations. Due to the large occupation
number, the small $x$ partons described by the EFT constitute a
color glass condensate \cite{iancu}.
 Since the classical fields of
the two nuclei are known, they provide the initial conditions for
the gluon fields produced in the nuclear collision, with the
particle production after the collision described by the
space-time evolution of classical gauge fields. The problem of
finding the evolution of the classical gauge fields in
nucleus-nucleus collisions has been addressed in lattice
simulations \cite{kraven}, allowing to compute the energy and the
number of produced gluons at central rapidities \cite{venprl}. On
the other hand,  in Ref. \cite{mueevo} Mueller has investigated
the approach to equilibrium in the McLerran-Venugopalan approach,
considering only elastic scattering and derived a Landau-type
equation for the single particle distributions $f(x,p)$. The
idealized initial condition for this transport equation is assumed
as given by
\begin{eqnarray}
f(x,p) = \frac{c}{\alpha_s N_c} \frac{1}{t} \delta(p_z)
\theta(Q_s^2 - p_t^2) \,\,, \label{dist}
\end{eqnarray}
where $c \approx 1.3$ is the parton liberation coefficient
accounting for the transformation of virtual partons in the
initial state to the on-shell partons in the final state. In Ref.
\cite{bjoraker}, the transport  equation has since been solved
numerically using the above initial condition and the approach to
equilibrium studied quantitatively, verifying that the
equilibration time and the initial temperature of the saturated
gluon plasma have a strong functional dependence on the initial
gluon saturation scale $Q_s$.
 Furthermore, Mueller has assumed that at early
times, in the maximum distance $1/q_{min}$ corresponding to the
minimal momentum transfer $q_{min}$ one has at most one
scattering.  This condition regularize the logarithmic collinear
divergence in the transport integral and provides an alternative
expression for the dynamical screening in the system of gluons
produced at central rapidities, which are  completely out of
equilibrium. One obtains \cite{mueevo}
\begin{eqnarray}
\mu^2 = (\frac{c\alpha_s N_c}{\pi})^{\frac{2}{3}}
\,\,\frac{Q_s^3}{(Q_s\,t)^{2/3}}\,\,. \label{debyesat}
\end{eqnarray}
As shown in Ref. \cite{bjoraker}, this expression for the
dynamical screening mass is larger than the equilibrium expression
Eq. (\ref{mubiro})  for  early times and is almost identical at
late times [The difference between the two expressions is $\approx
0.05$ GeV for $t\approx 0.04 fm$ and is approximately zero for $t
= 1 fm$]. As the heavy quark production and the formation of bound
states are predicted to occur in the initial stage of the
deconfined phase, we will use the expression (\ref{debyesat}) for
the dynamical screening mass in our  analyzes, which implies an
upper limit for the predictions of quarkonium suppression.

Now let us consider a central collision in a nucleus-nucleus
collision, which results in the formation of a quark-gluon plasma.
Consider the central slice ($y=0$) of a collision which leads to
the production of a heavy quark pair from initial fusion and to a
QGP. During the initial interaction period, heavy quarks will be
produced through gluon fusion and quark-antiquark annihilation.
Like gluon and light quark production, heavy quark production
through the initial fusion is very sensitive to the parton
distributions inside the nuclei. Although the cross sections for
heavy quark production in RHIC and LHC are strongly modified by
the presence of the  high density effects \cite{ayavic,ayavic2},
the quarkonium  production rate should be large enough for any
suppression to be observable. The charm quarks will be produced on
the order of a time scale $1/2m_c \approx 0.07 fm$, while the
bottom quarks will similarly be produced over a time of $\approx
0.02 fm$. Thus immediately upon production, these quarks will find
themselves in a deconfined matter which is highly out of
equilibrium and rapidly evolving towards kinetic equilibration. In
the plasma rest frame, the $Q\overline{Q}$ forms a bound state in
the time $t_F = \gamma \tau_F$, where $\gamma =
\sqrt{1+(p_T/M)^2}$ ($M$ is the onium mass) and the formation
times, $\tau_F$, are $\tau_F^{\Upsilon} = 0.76 fm$ and
$\tau_F^{J/\Psi} = 0.89 fm$. Here we will concentrate in the small
$p_T$  region, where $t_F \approx \tau_F$. The color charge of the
heavy  quark will be subject to screening due to the presence of
the  saturated gluon plasma. If the dynamical screening mass $\mu$
[Eq. (\ref{debyesat})] is larger than the critical screening mass
$\mu_i^{diss}$ [Table 1],  the $Q$ and $\overline{Q}$ cannot form
a bound state and the $Q\overline{Q}$ system will dissociate into
a separate $Q$ and $\overline{Q}$, which subsequently hadronize by
combining with light quarks or antiquarks to emerge as "open
flavors" mesons. We will concentrate our analyzes in the
$\Upsilon$ suppression at LHC energies, since at current energies,
the situation of $J/\psi $ suppression is rather ambiguous because
the bound state can also break up through interactions with
nucleons and comoving hadrons, i.e., QGP production has not been
proved to be the unique explanation of the observed suppression
even though an increased density of secondary production is
needed. Because the $\Upsilon $ is much smaller than
$c\overline{c}$, a much higher dynamical screening mass  is
necessary to dissociate the $\Upsilon $, providing a valuable tool
to determine the initial state of the system and the
characteristics of the plasma \cite{gunion}. Moreover, we will
consider the dynamical evolution of the screening mass for  $ t
> t_0$, where $t_0$ is inferior limit for the application of the transport theory.
For the LHC energies which we are interested $t_0 \le 0.18 fm$
\cite{bjoraker}.

In Fig. \ref{fig1} we present the time dependence of the dynamical
screening mass as predicted in a saturated gluon plasma for some
values of the initial gluon saturation scale. We can see that our
predictions are very sensitive  on the saturation scale.
Considering that the coupling constant is constant and equal to
0.3, we have that for small values of the saturation scale ($Q_s
\approx 1.0$ GeV) our results predict that the meson $\Upsilon$ is
not dissociated, causing it to escape unscathed. For medium values
of $Q_s$ ($\approx 2.0$ GeV) the dynamical screening mass is
larger than the dissociation mass only for the early stage of
evolution, where the formation of a bound state is unexpected. A
similar scenario is obtained if we consider the values of $c$
(=1.23), $\alpha_s$ (=0.6) and $Q_s$ (=1.41) derived in Ref.
\cite{nardi}, where the recent data of the PHOBOS collaboration
for the charged multiplicity distribution were analyzed
considering the saturation scenario. Therefore, if the saturation
scenario is valid at RHIC with the typical values for $\alpha_s$
and $Q_s$ extract from the multiplicity distribution we does not
expect the $\Upsilon$ suppression for the current energies in this
collider ($\sqrt{s} = 130 GeV$).  Only for large values of the
saturation scale we predict the $\Upsilon$ suppression, since by
the time that the upsilon is formed the dynamical screening mass
is above of $\mu _\Upsilon ^{diss}$, required to inhibit its
formation.

In Fig. \ref{fig2} we present the dependence of the dynamical
screening mass in the the coupling constant for two typical values
of the saturation  scale. We analyze the behavior of the dynamical
screening mass in the interval of times between the inferior limit
for the application of the transport theory ($t \approx 0.2 $ fm)
and the characteristic value of formation time ($t \approx 0.8$
fm). We verify that if the plasma is characterized for a small
value of $Q_s$ , the suppression of the meson $\Upsilon$ is only
expected for early times of evolution and large values of the
coupling constant, where the perturbative calculations becomes
questionable. For large values of the saturation scale, the
$\Upsilon$ suppression is predicted for $\alpha_s \ge 0.3$ in the
initial stage of the kinetic evolution. However, at late stages
the $\Upsilon$ dissociation  is not expected. For comparison, we
also present in Fig. \ref{fig2} the dissociation mass of the
$J/\Psi$ meson. We verify that in this case the complete
suppression is expected to occur in the presence of a saturated
gluon plasma.

In Fig. \ref{fig3} we present the dependence of the dynamical
screening mass in the saturation scale for a typical value of the
coupling constant. We verify that the dissociation of the
$\Upsilon$ must occur only if the saturated gluon plasma is
characterized by large values of the momentum saturation scale. In
other words, the presence of $\Upsilon$ suppression indicates that
if a saturation scenario is valid, the initial nuclear wave
function are characterized for a bulk momentum scale $Q_s \ge 1.5$
GeV. Therefore, the $\Upsilon$ suppression may be an indirect
probe of the nuclear wave functions, and consequently, from the
Color Glass Condensate, allowing to constraint the initial gluon
saturation scale.

Some comments are in order. In this Letter we have assumed that
the saturation scale is in the range 1-3 GeV at the RHIC and LHC
energies, and derive the respective pattern of $\Upsilon$
suppression. A more precise constraint of the saturation scale
value can be obtained if we consider the $eA$ processes, which
could be analyzed in the future $eA$ colliders at RHIC and HERA.
In particular, the analyzes of the slope of the nuclear structure
function \cite{vicslope} and the ratio between the transverse and
longitudinal structure function \cite{mcllev} are potential
observables to determinate the value of $Q_s$. Our results
demonstrate that this independent constraint of the saturation
scale is fundamental for the description of the observables in
$AA$ process.

Lets compare our results with other predictions of $\Upsilon$
suppression. In Ref. \cite{satzdin} the expression (\ref{mubiro})
has been used to analyze the  evolution of the dynamical
screening, as well as the onset of quarkonium suppression,
considering initial conditions provided by the parton cascade
model. One of the main conclusions of that analyzes is that the
$\Upsilon$ suppression is only expected at LHC energies.
Distinctly from our results, that predicts that the suppression
must occur only for the quarkonia produced in the initial stages
and at large values of the saturation scale,  these authors
predicts the complete suppression, since the screening mass is
ever above of the dissociation mass in the dynamical evolution.
Such result is associated to the initial conditions used as input
in the calculations, which does not consider a saturation
criterium in the nuclear wave functions. Similar situation occur
if we compare our results with that obtained in Ref. \cite{patra},
where the quarkonium suppression in an equilibrating quark-gluon
plasma was evaluated considering two mechanisms: the Debye
screening and the gluonic dissociation, concluding that the Debye
screening is not effective for $\Upsilon$ suppression at the LHC
energy. This result derives from the self screened parton cascade
model, which  leads to small Debye mass and implies that at the
formation time of the $\Upsilon$,   the Debye mass drops below the
value of $\mu _\Upsilon ^{diss}$, required to inhibit its
formation, causing it to escape unscathed. Therefore, the main
difference between the predictions for $\Upsilon$ suppression is
associated to the use of  distinct initial condition for the
dynamical evolution. Consequently, independent searches of the
gluon saturation of the  nuclear wave function are required to
discriminate between the distinct approaches.

In brief, we have seen that if the saturation scenario is valid
for RHIC and LHC, with the  nuclear wave functions characterized
by a bulk momentum scale $Q_s$, the pattern of   quarkonium
suppression is also $Q_s$ dependent. We have verified that the
$\Upsilon$ suppression is only expected at large values of the
saturation scale, allow us to constraint the values of this scale.
In other words, the presence of $\Upsilon$ suppression is an
indicative of the presence of a saturated gluon plasma,
characterized by a large value of the initial gluon saturation
scale in the nuclear wave functions. Of course,  more quantitative
analyzes can be made if this scale was derived in an independent
process.

\section*{Acknowledgments}
The author acknowledge helpful discussions with A. Ayala
(IFM-UFPel), M. B. Gay Ducati (IF-UFRGS) and the members of the
Phenomenology Group at Physics Institute - UFRGS. This work was
partially financed by CNPq  and FAPERGS, BRAZIL.

\newpage

\section*{Tables}

\begin{table}[h]
\begin{center}
\begin{tabular} {||l||l|l|l|l|l|l||}
\hline \hline State & $J/\psi $ & $\psi ^{\prime }$ & $\chi _c$ &
$\Upsilon $ & $\Upsilon
^{\prime }$ & $\chi _b$ \\
\hline $\mu _i^{diss}$ (GeV) & 0.70 & 0.36 & 0.34 & 1.57 & 0.67 &
0.56 \\\hline \hline
\end{tabular}
\end{center}
\label{tab1}
\caption{The dissociation screening mass for
quarkonia.}
\end{table}

\newpage

\section*{Figure Captions}

\begin{figure}[t]
\centerline{\psfig{file=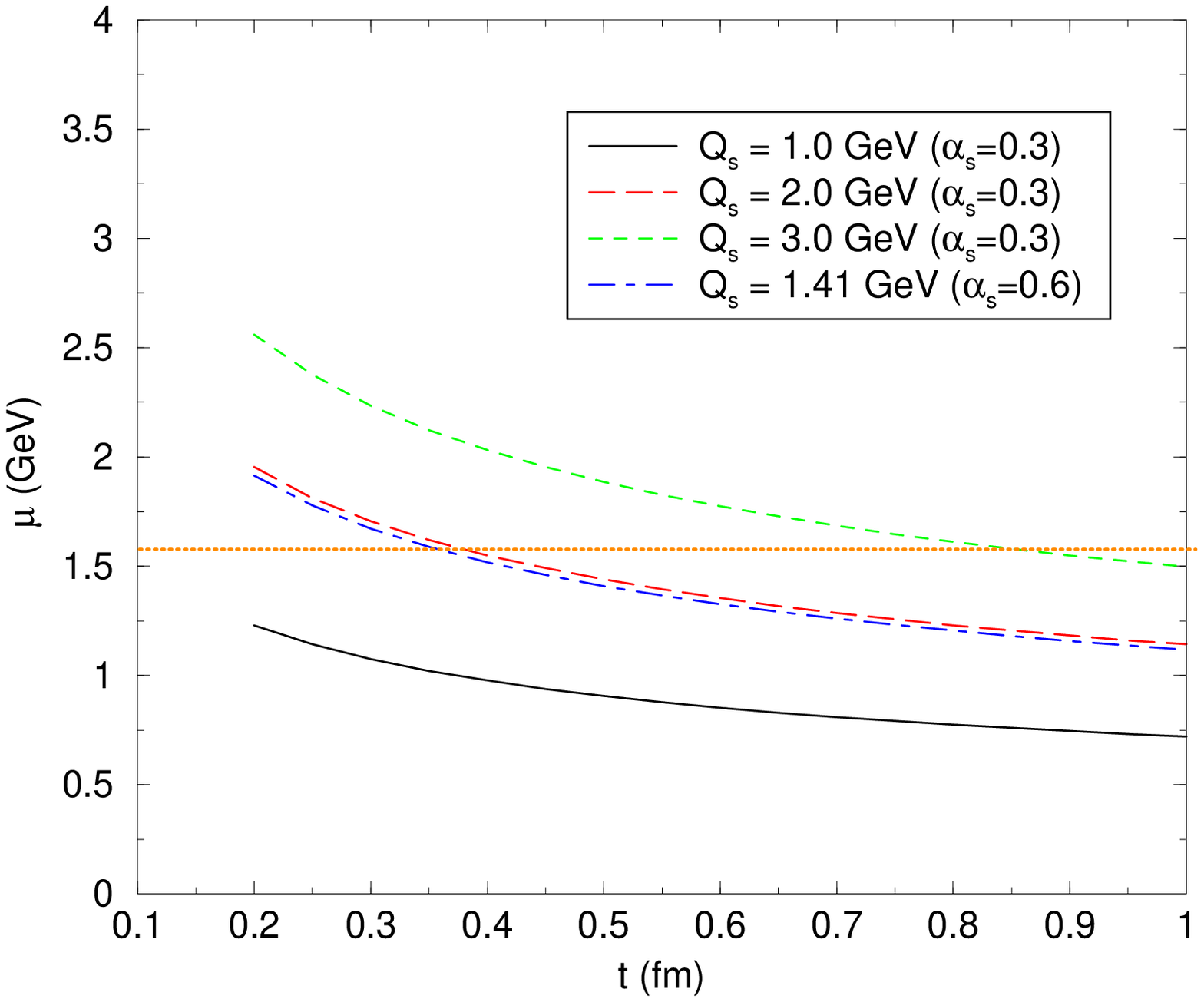,width=150mm}} \caption{}
\label{fig1}
\end{figure}

\vspace{1.0cm} Fig. \ref{fig1}: The time dependence of the
screening mass in a saturated gluon plasma for some typical values
of the saturation scale $Q_s$ and the coupling constant
$\alpha_s$. The horizontal line characterize the dissociation
screening mass for the meson $\Upsilon$.

\begin{figure}[t]
\begin{tabular}{c c}
\psfig{file=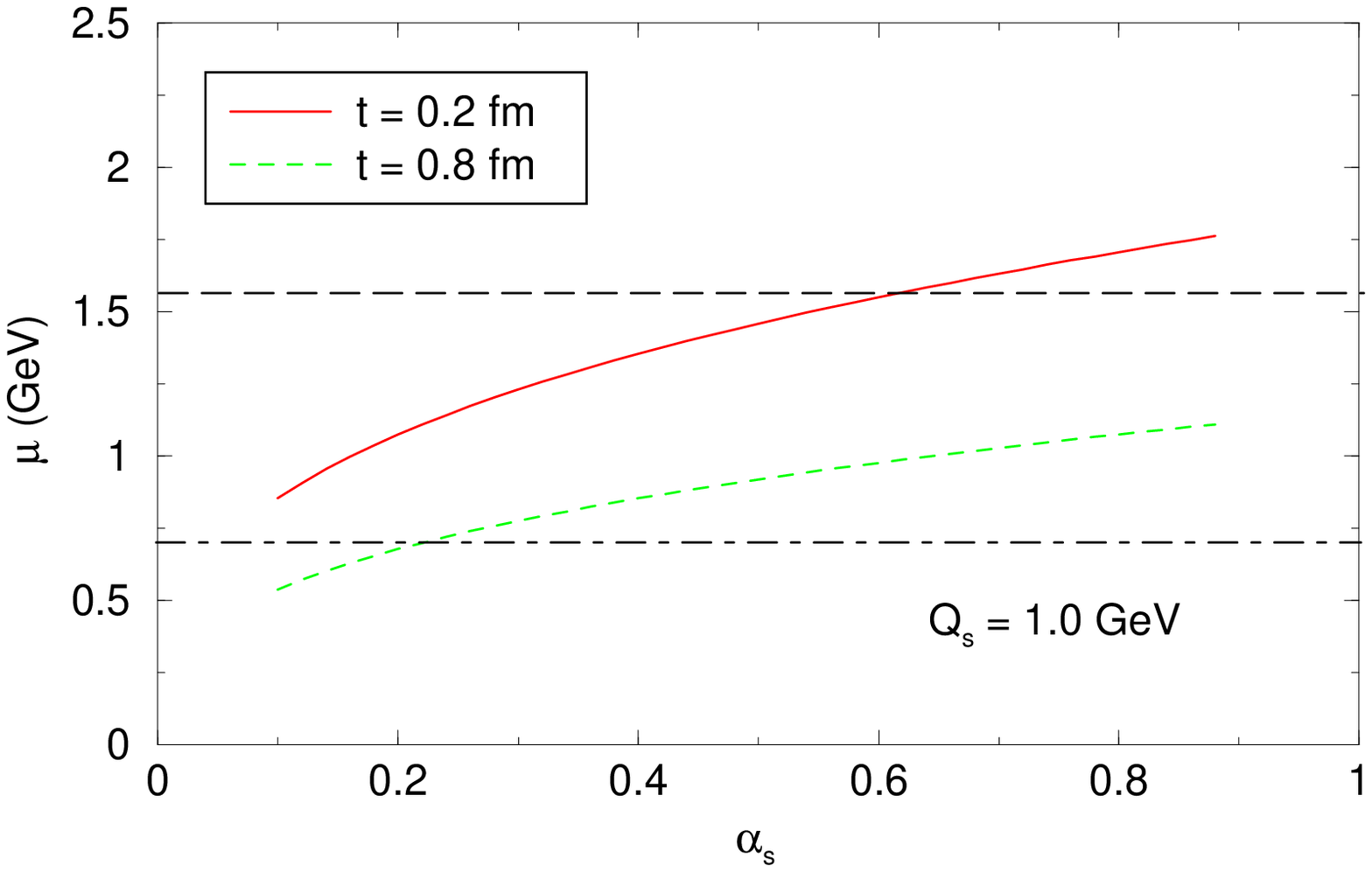,width=100mm} \\
\psfig{file=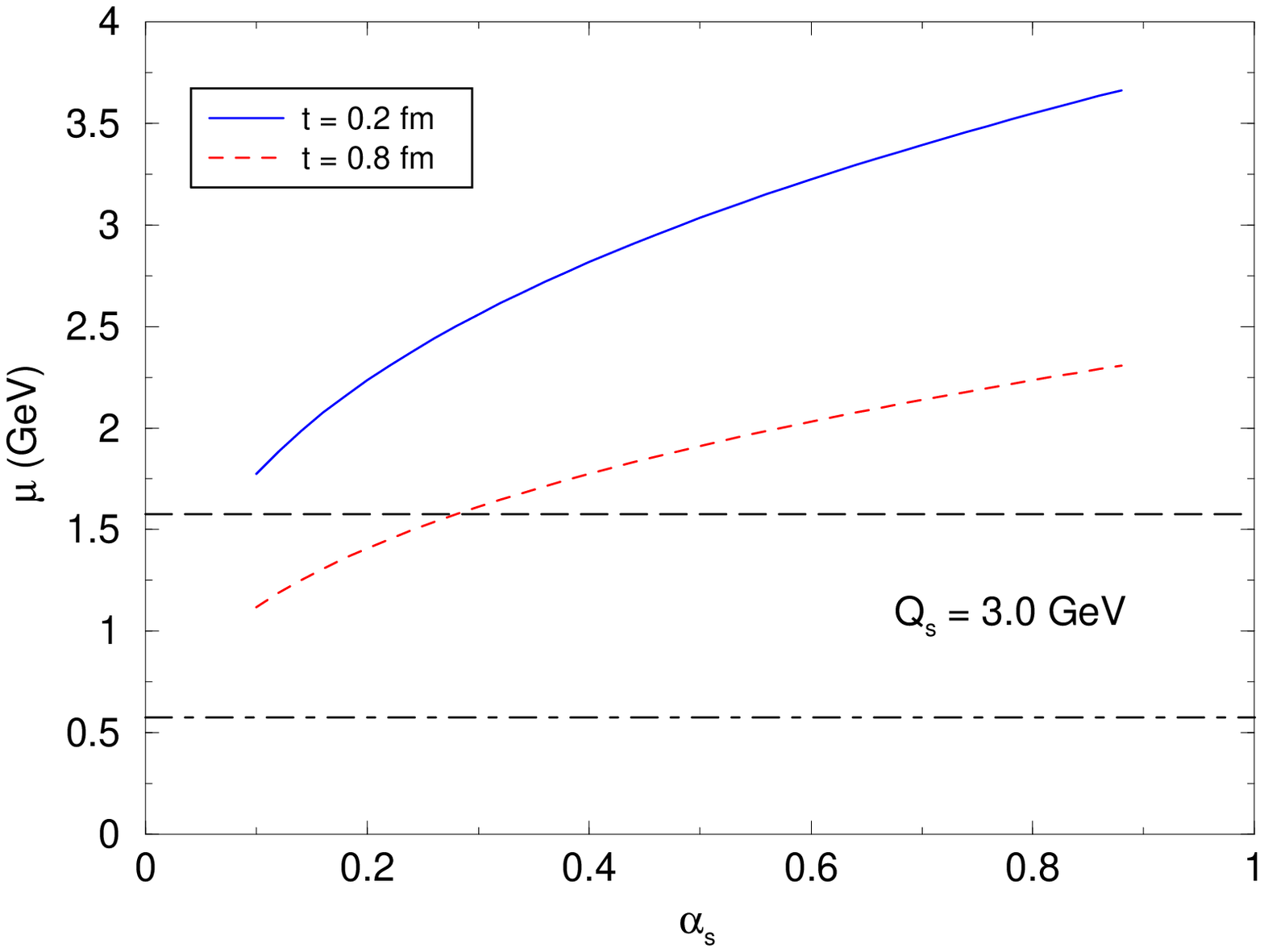,width=100mm}
\end{tabular}
\caption{}
\label{fig2}
\end{figure}

\vspace{1.0cm} Fig. \ref{fig2}: The dependence of the screening
mass in the coupling constant for two instants in the saturated
plasma evolution. The long dashed and dotted-dashed curves
characterize the dissociation screening mass for the mesons
$\Upsilon$ and $J/\Psi$, respectively.

\begin{figure}[t]
\centerline{\psfig{file=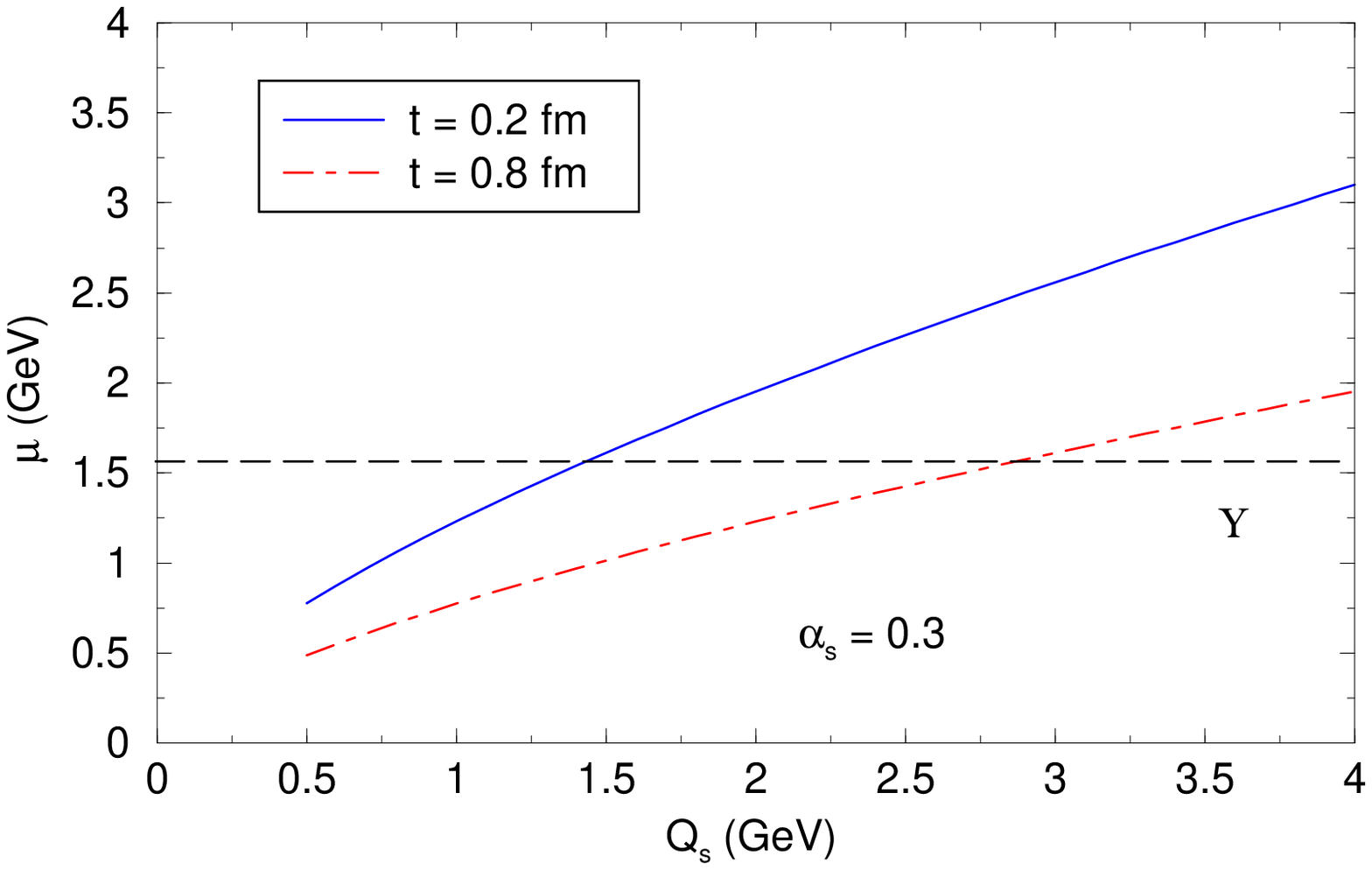,width=150mm}} \caption{}
\label{fig3}
\end{figure}

\vspace{1.0cm} Fig. \ref{fig3}: The dependence of the screening
mass in the saturation scale for two instants in the saturated
plasma evolution. The long dashed  curve characterize the
dissociation screening mass for the meson $\Upsilon$.

\newpage

\newpage

\newpage

\end{document}